\documentclass[table]{ccjnl}
\pdfoutput=1
\title{Meta Reinforcement Learning for Fast Spectrum Sharing in Vehicular Networks}
\author{
Kai Huang\inst{1}, Le Liang\inst{1,2,*}, Shi Jin\inst{1,*}, Geoffrey Ye Li\inst{3}
\corinfo{$\left\{\textrm{lliang, jinshi}\right\}$@seu.edu.cn}
}
\receiveddate{April 4, 2023}
\reviseddate{July 10, 2023}
\Editor{Honggang Zhang}

\address[1]{National Mobile Communications Research Laboratory and Frontiers Science Center for Mobile Information Communication and Security, Southeast University, Nanjing, 210096, China}
\address[2]{Purple Mountain Laboratories, Nanjing, 211111, China}
\address[3]{Department of Electrical and Electronic Engineering, Imperial College London, London SW7 2AZ, U.K.}

\begin{document}
\maketitle

\begin{abstract}
In this paper, we investigate the problem of fast spectrum sharing in vehicle-to-everything communication. In order to improve the spectrum efficiency of the whole system, the spectrum of vehicle-to-infrastructure links is reused by vehicle-to-vehicle links. To this end, we model it as a problem of deep reinforcement learning and tackle it with proximal policy optimization. A considerable number of interactions are often required for training an agent with good performance, so simulation-based training is commonly used in communication networks. Nevertheless, severe performance degradation may occur when the agent is directly deployed in the real world, even though it can perform well on the simulator, due to the reality gap between the simulation and the real environments. To address this issue, we make preliminary efforts by proposing an algorithm based on meta reinforcement learning. This algorithm enables the agent to rapidly adapt to a new task with the knowledge extracted from similar tasks, leading to fewer interactions and less training time. Numerical results show that our method achieves near-optimal performance and exhibits rapid convergence.
\end{abstract}

\section{Introduction}\label{Introduction}
Vehicle-to-everything (V2X) communication has been recognized as a critical technology to enhance connected and intelligent transportation services in many aspects \cite{1}, such as road safety and traffic efficiency. The 3rd Generation Partnership Project (3GPP) has currently unrolled the support of V2X services in long-term evolution (LTE) \cite{2} and 5G new radio (NR) \cite{3}. These cellular vehicular communication technologies, or C-V2X, have gained increasing attention from both academia and industry. 
\subsection{Motivation}
This paper mainly considers the design of spectrum sharing and power selection in vehicular networks. The vehicle-to-infrastructure (V2I) link and vehicle-to-vehicle (V2V) link are two crucial communication links that enable V2X communication. The V2I links facilitate communication between vehicles and the base station (BS) or roadside unit (RSU), with a primary focus on high data rate applications. On the other hand, the V2V links establish direct communication links between neighboring vehicles and are responsible for the transmission of safety-critical messages. These safety messages usually consist of information such as vehicle position, velocity, driving direction, etc., which enhance the awareness of other vehicles. 

This paper considers Mode 4 in 3GPP LTE-V2X or Mode 1 in NR-V2X, in which vehicles have a pool of radio resources to select autonomously for communication via the V2V links \cite{4}. Because of the limited radio resource in a cellular-based vehicular network, the V2V links have to share spectrum with the V2I links in order to improve the spectrum efficiency. However, resource allocation in vehicular networks poses a significant challenge due to their inherent dynamics and the varying quality-of-service (QoS) requirements. In order to satisfy these divergent requirements, a well-conceived resource allocation strategy is necessary to effectively manage interference and facilitate the coexistence of these two links. To simplify this problem, we make the same assumption as \cite{5,6} that each V2I link has preoccupied an orthogonal spectrum with fixed transmission power. Therefore, the optimization problem is left for the proper design of the V2V links.

\subsection{Related Works}
Numerous research works have used traditional optimization methods to address the issue above. For instance, in \cite{7}, the signaling overhead is reduced using slow fading component of channel state information (CSI), maximizing the sum ergodic capacity of V2I links and ensuring the reliability of V2V links simultaneously. Additionally, \cite{8} proposes an algorithm based on Lagrange dual decomposition to solve the resource allocation problem within reasonable complexity. Moreover, \cite{9} tackles the spectrum sharing problem in a general form of V2X communication with an approach using graph theory. Despite the great performance achieved by these approaches, several obstacles remain. The first one is that global CSI is often required, so these methods are usually executed in a centralized manner, leading to significant network signaling overhead for CSI acquisition. Secondly, the computational complexity of these methods is usually unacceptable for practical implementation. In addition, these methods tend to disregard the dynamics of the channel evolution, leading to bad performance in tackling sequential decision making problems.

Reinforcement learning (RL) holds the potential in handling sequential decision making problems and therefore it can be adopted to address the challenges mentioned above. Nowadays, RL has been widely leveraged to address many challenging problems in wireless communications, such as beam selection and precoding in MIMO systems \cite{10}, phase shift design in reconfigurable intelligent surface (RIS) assisted systems \cite{11} and wireless resource allocation \cite{12}. Wireless resource allocation is particularly shaped by RL as it enables the training for objectives that are difficult to model due to the flexible design of the reward function. A more attractive point is that RL provides a distributed structure for the resource allocation scheme so that the overhead and delay in the system can be efficiently decreased. In \cite{13}, deep RL has been leveraged to solve a dynamic spectrum access problem, where the user needs to choose a good channel for transmission with the history of previous choices and results. Numerical results show that RL-based method can closely approach the genie-aided myopic policy, which is known as the optimal solution for this problem. Besides, for power allocation problems, where multiple communication links share a single spectrum band aiming to maximize generic weighted sum rate, a model-free deep RL-based scheme has been developed in \cite{14}. They show that this scheme can track the channel evolution and be executed in a distributed manner, which outperforms weighted minimum mean-squared error (WMMSE) or fractional programming (FP) algorithm. For the joint spectrum and power allocation problem, a decentralized single-agent DQN-based approach is developed in \cite{5} for the unicast and broadcast scenarios in V2X communication to select the optimal sub-band and transmission power. In addition, \cite{6} utilizes a multi-agent RL-based method, building upon the work in \cite{5}, to enable cooperative learning among multiple V2V agents, thereby enhancing overall system performance.

Although RL is a promising solution for addressing the resource allocation problem in V2X communication, several challenging issues remain unresolved. Firstly, a simulation environment is typically needed as training a well-performing policy necessitates extensive interactions with the environment. When the real and simulation environments are identical, the policy trained on simulators can attain desirable performance. However, differences between these two environments, referred to as the reality gap, may result in reduced efficacy of a well-trained agent in the simulation environments when it is deployed in a live network. Additionally, training a policy directly in the real environment is often infeasible due to the requirement of an extensive number of interactions. Besides, an immature policy is improper for collecting the experience data as it may occasionally lead to catastrophic consequences in practice. Recently, there are several works trying to solve this challenge. For example, a distributed user scheduling and downlink power control scheme in multi-cell wireless networks is proposed in \cite{15}. As the users of each access point vary between the simulation and the real environment, a fixed number of users are sorted according to the proportional fairness criterion to improve the scalability, whose information will be the input of the neural network. However, these methods are usually scenario-specific, so they cannot be adopted to tackle other kinds of problems easily.

Motivated by the human ability to learn new tasks rapidly with experience in similar tasks, the generalization of the policies can be improved via meta learning. Recently, meta learning is mainly divided into two categories, one is gradient-based meta learning and the other is recurrent neural network (RNN) based meta learning. Gradient-based meta learning aims to obtain an initialization, which enables the policy to adapt to a new task with minimal gradient descents, by training across many similar tasks. Model-agnostic meta-learning (MAML) is a prominent algorithm of this kind of meta learing that has gained significant popularity due to its effectiveness in regression and deep RL tasks \cite{16}.  In \cite{17}, the computational complexity of MAML is reduced by several similar meta learning algorithms that can achieve comparable performance, including first-order MAML and Reptile. Besides, there are some works that utilize the RNN to extract useful information from the tasks, and in some cases, they have achieved much better performance than gradient-based meta learning, such as RL$^2$ \cite{18}, PEARL \cite{19}, VariBAD \cite{20} and so on. 

Based on meta learning and RL, meta reinforcement learning (meta-RL) is an amalgamation of them, which is widely employed to address various problems in wireless communication, including trajectory design of unmanned aerial vehicles \cite{21,22}, channel estimation using pilot blocks \cite{23}, adaptive bit-rate decision for heterogeneous network conditions \cite{24}, task offloading in mobile edge computing \cite{25}, load balancing for multi-band downlink cellular networks \cite{26}, and reliable communication for terahertz or visible light communication networks \cite{27}. They have highlighted the potential of meta-RL to enhance the generalizability and versatility of RL in wireless communication.
\subsection{Contributions}
In this paper, to deal with the problem of reality gap in the vehicular networks, we adopt meta-RL to design a resource allocation scheme, with which we can obtain an initialization with good generalization for our policy. The main contributions of this paper are summarized as follows:
\begin{itemize}
	\item[1.] We design a joint optimization scheme for spectrum sharing and power selection in vehicular networks based on single-agent proximal policy optimization (PPO) method.
	\item[2.] We propose an algorithm based on meta-RL to obtain an efficient spectrum sharing scheme more rapidly in an unseen environment. The whole algorithm contains the training stage and the adaptation stage, and the training stage consists of two loops. We use the PPO method to get the parameters for specific environments in the inner loop. We choose Reptile rather than the MAML method to obtain meta parameters in the outer loop to decrease the computational complexity. 
	\item[3.] Simulations are presented to demonstrate the performance of our proposed meta-RL algorithm. We demonstrate our algorithm is capable of acquiring a proficient initialization, with which the policy can adapt to the new tasks within merely a limited number of interactions. Furthermore, our scheme can achieve desirable performance and converge rapidly. Moreover, we also investigate how the total number of tasks in the meta-training stage and the difference between the new environment and training environments affect the performance of the proposed method.
\end{itemize}
\subsection{Organization}
The rest of the paper is organized as follows. In Section ~\ref{model}, we introduce the system model of a cellular-based uplink vehicular network and formulate the optimization problem of spectrum sharing and power selection. A meta-RL based resource allocation algorithm is proposed in Section ~\ref{algorithm} to solve this problem. Simulation results are provided in Section ~\ref{simulation} and the conclusion is drawn in Section ~\ref{conclusion}.

\section{System Model and Problem Formulation}\label{model}
\begin{figure}[htbp]
    \centering
    \includegraphics[width=0.45\textwidth]{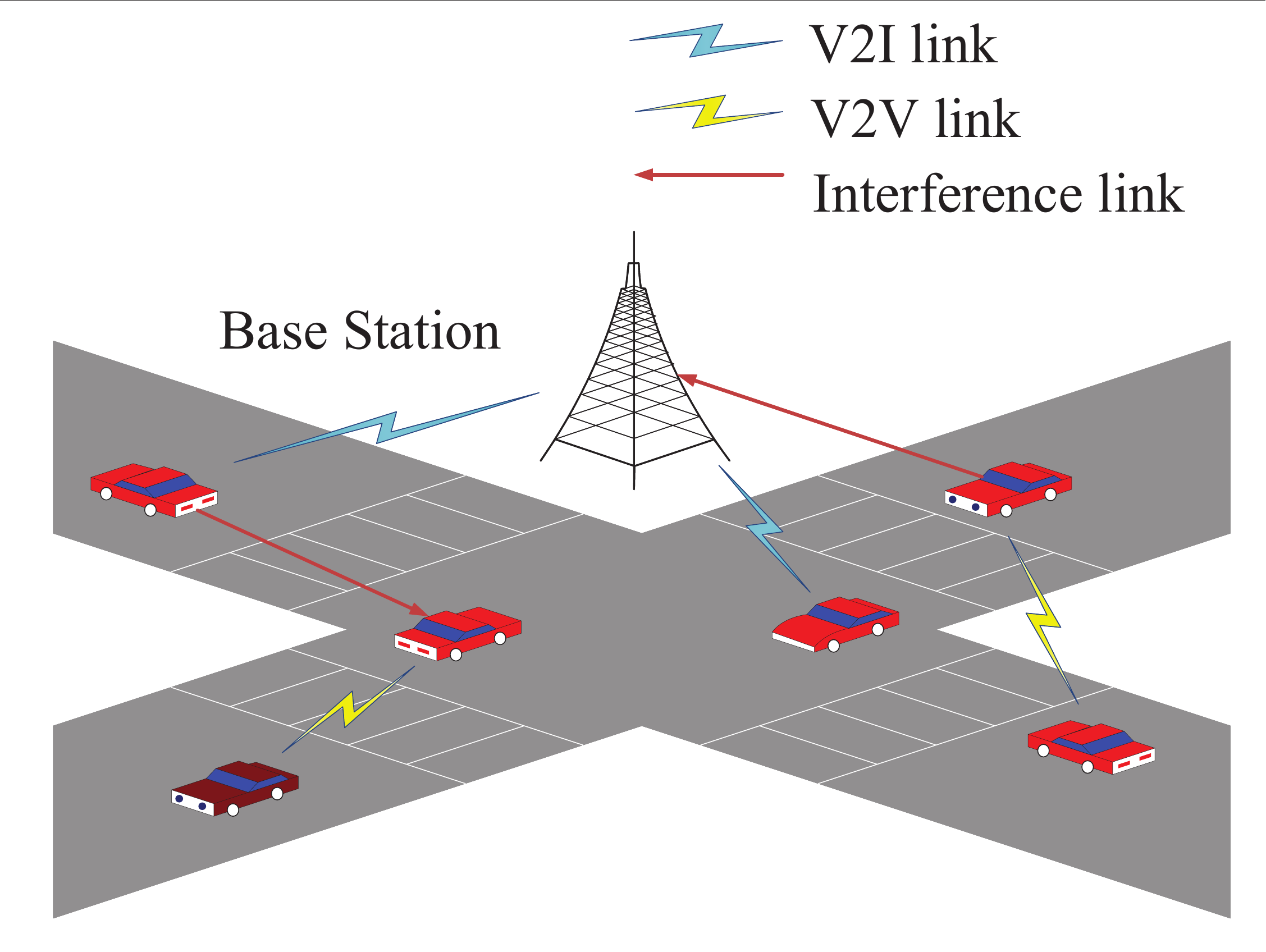}
    \caption{An illustrative structure of a cellular-based vehicular networks.}
	\label{fig1}
\end{figure}

A cellular-based vehicular network is considered in this paper, of which the structure is shown in Fig. \ref{fig1}. There are $A$ V2I links and $N$ V2V links in the network, denoted by $\mathcal{A}=\{1,\dots,A\}$ and $\mathcal{N}=\{1,\dots,N\}$, respectively. Only the uplink process of V2I links is considered, and each V2V link autonomously selects the spectrum to access and the power for transmission. It is assumed that there are $A$ orthogonal sub-bands in the system, and each sub-band corresponds to a V2I link, of which the transmission power is fixed as $P_{a}^{\textrm{V2I}}$. Besides, the sub-bands are reused by the V2V links in order to improve the spectrum efficiency.

The channel gain of the $a$th sub-band from the $k$th V2V transmitter to the $n$th V2V receiver is denoted as: 
\begin{equation}
   g_{k,n}[a]=\alpha_{k,n} h_{k,n}[a],  
\end{equation}
where $\alpha_{k,n}$ captures the slow-fading effect that is independent with frequency, with path loss and shadowing included, and $h_{k,n}[a]$ is the frequency-relative fast-fading component. Similarly, the channel gain of the $a$th sub-band for the $a$th V2I link is defined as $\overline{g}_{a,B}[a]$ while for the $n$th V2V link it is $g_n[a]$. There are two kinds of interfering channels: the channels from the V2V transmitters to the BS and the channels from the V2I transmitters to the V2V receivers. And their channel gain can be similarly denoted as $g_{n,B}[a]$ and $\overline{g}_{a,n}[a]$, respectively. Therefore, the received signal-to-interference-plus-noise ratio (SINR) for the $a$th uplink V2I link over the $a$th sub-band can be obtained through
\begin{equation}
\gamma _{a}^{\textrm{V2I}}[a] = \frac {P_{a}^{\textrm{V2I}} \overline {g}_{a,B}[a]}{\sigma ^{2} + \sum \limits _{n}P_{n}^{\textrm{V2V}}[a] g_{n,B}[a]},
\label{eq1}
\end{equation}
and the SINR for the $n$th V2V link over the $a$th sub-band is
\begin{equation} 
\gamma _{n}^{\textrm{V2V}}[a] = \frac {P_{n}^{\textrm{V2V}}[a] g_{n}[a]}{\sigma ^{2} + I_{n}[a]},
\label{eq2}
\end{equation}
where 
\begin{equation}
I_n[a]=P_{a}^{\textrm{V2I}} \overline {g}_{a,n}[a] + \sum \limits _{k \ne n}P_{k}^{\textrm{V2V}}[a] g_{k,n}[a],
\label{eq3}
\end{equation} 
denotes interference power, $\sigma^{2}$ is the power of noise and $P_{n}^{\textrm{V2V}}[a]$ denotes the power of the $n$th V2V link for transmission over the $a$th sub-band. $P_n^{\textrm{V2V}}[a]>0$ means that the $a$th sub-band is occupied by the $n$th V2V link, and $P_n^{\textrm{V2V}}[a]=0$ otherwise\inst{1}\let\thefootnote\relax\footnotetext{\inst{1}Please note that we can equivalently use binary variables $\rho_n[a]\in\left\{0,1\right\}$ to indicate spectrum allocation and continuous variables $P_{n}^{\textrm{V2V}}[a]$ for power control, as is the case in \cite{6,28}.}.

As the V2I links aim to support the applications with high data rate in the vehicular network, the sum rate of V2I links, that is denoted as $\sum\limits_{a}C_{a}[a]$, should be maximized, where 
\begin{equation} 
C_{a}[a]=W\textrm{log}(1+\gamma _{a}^{\textrm{V2I}}[a]).
\end{equation}

The V2V links are responsible for transmitting safety-critical information with high reliability so they are designed to improve the success probability of transmission, which in this paper is modeled in mathematical form as:
\begin{equation} 
\text {Pr}\left \{{ \sum _{t=1}^{T}\sum \limits _{a=1}^{A} C_{n}^{\textrm{V2V}}[a,t] \ge B/\Delta T}\right \},\quad n\in \mathcal {N},
\label{eq4}
\end{equation}
where 
\begin{equation}
C_{n}^{\textrm{V2V}}[a,t]=W\textrm{log}(1+\gamma _{n}^{\textrm{V2V}}[a,t]),
\end{equation}
denotes the maximum achievable rate of the $a$th sub-band for the $n$th V2V link in time slot $t$. The channel coherence time is $\Delta T$ and the size of payload to be transmitted by the V2V links is denoted as $B$ in bits. 

Therefore, our spectrum and power allocation problem is to design the transmission power $P_n^{\textrm{V2V}}[a]$ of the V2V links, for all $n\in\mathcal{N}$, $a\in\mathcal{A}$, aiming to maximize the sum rate of V2I links and the success probability of transmission of V2V links simultaneously.

\nocite{29}
\section{Meta-RL Based Spectrum Sharing Design}\label{algorithm}
The proposed spectrum sharing problem is actually a sequential decision making problem, and we can model it as a Markov decision process. As introduced in Section \ref{Introduction}, deep RL can be utilized to solve this kind of problems efficiently. Nevertheless, it is also mentioned that an environment simulator is often required for training a well-performing policy. Since the simulator has some inevitable differences from the live network, an agent may suffer severe performance degradation when used in practice, despite its optimal performance achieved in the simulation environment.  

In this paper, a meta-RL based algorithm is proposed to solve this problem. Meta-RL aims to obtain the parameters through training the agent with a large number of tasks, which can be used as the initialization for the policy networks to help the agent adapt faster when deployed in the real world. We can generate many similar tasks by adjusting the values of the key factors of the environments for V2X communication. The task mentioned in this paper refers to obtaining an efficient policy for spectrum sharing in a vehicular environment, of which the key factors are known. 

Our algorithm consists of two stages: the meta-training stage, which obtains the meta parameters through a large number of similar tasks, and the adaptation stage, which commences by initializing the policy network with the meta parameters and then adapts for several episodes in the new environment. The meta-training stage includes two loops: the inner and outer loops. The inner loop updates the policy parameters with the experience data from a specific environment. The outer loop amalgamates the updated parameters from different inner loops and updates the meta parameters. Next, we introduce the details of our design.

\subsection{Design of the Inner Loop}\label{inner}
In short, the inner loop solves a typical RL problem. Here, each V2V link in the network is regarded as an agent.
\begin{itemize}
    \item {\textbf{The observation space}}: Local CSI, received interference power, remaining V2V payload and time for transmission are contained in the observation space. The state of the $n$th V2V link in time slot $t$ is denoted as
    \begin{equation}
        S_t^n=\{\left\{G_n[a]\right\}_{a\in\mathcal{A}}, \left\{I_n[a]\right\}_{a\in\mathcal{A}}, B_n, T_n\},
        \label{eq5}
    \end{equation}
    with $G_n[a]=\left\{g_n[a],g_{k,n}[a],g_{n,B}[a],\overline{g}_{a,n}[a]\right\}$.
    \item {\textbf{The action space}}: In this paper, the V2V link can only choose its transmission power among four discrete values, [23, 15, 5, -100] dBm, for the ease of training and practical circuit implementation. It can also be generalized to a continuous space. Since the majority of meta-RL algorithms are model-agnostic, they remain effective when the neural network architecture changes. Here, each action is a combination of spectrum and transmission power selection with the dimension of $M\times 4$.
    \item {\textbf{The reward function}}: It is designed as the weighted sum rate of these two kinds of links:
    \begin{equation}
        r_t=\lambda_a\sum_{a}C_a[a,t]+\lambda_n\sum_{n}R_n(t),
        \label{eq6}
    \end{equation}
    where $\lambda_a$ and $\lambda_n$ are positive weights. $R_n(t)$ is the reward related with the V2V rate in time slot $t$:
    \begin{equation} 
        R_{n}(t) = \begin{cases} \displaystyle \sum \limits _{a=1}^{A} C_{n}^{\textrm{V2V}}[a,t], & \text {if $B_{n} > 0$},\\ \upsilon, & \text {otherwise}. \end{cases}
        \label{eq7}
    \end{equation}
    In order to encourage V2V links to finish their deliveries faster, the value of $\upsilon$ is greater compared with the largest possible V2V rate. We can approximate it by rolling out several episodes and selecting the largest rate among these time slots. 
\end{itemize}

\begin{algorithm}[htbp]
	\caption{\large Meta-RL based Spectrum Sharing for V2X communication: Meta-training Stage}
	\label{alg1}
	\begin{algorithmic}[1]
		\Require
		a meta training set: $\left\{\mathcal{T}\right\}\sim p(\mathcal{T})$\newline
		the step sizes for two loops: $(\mu,\epsilon)$
		\Ensure
		meta parameters: $\psi$ (actor), $\omega$ (critic)
		\State Initialize the meta parameters $\psi$ and $\omega$ randomly.
		\For{outer loop $i \in [1,N_{\textrm{out}}]$}         
		\State Sample $N_{\textrm{task}}$ tasks from the training set $\left\{\mathcal{T}\right\}$.
		\For{task $j\in[1,N_{\textrm{task}}]$}
		\State $\psi_j\leftarrow\psi$, $\omega_j\leftarrow\omega$. 
		\State Empty the experience buffer.
		\For{inner loop $k \in[1,N_{\textrm{in}}]$}
		\State Collect experience data $\mathcal{D}_k$ with $\pi_{\psi_j}$. 
		\State Store $\mathcal{D}_k$ to the $j$th experience buffer.
		\For {gradient step $s\in[1,N_U]$}
		\State Sample data from the $j$th buffer.
		\State Update parameters:
        \State $\psi_j\leftarrow\psi_j+\mu\nabla_{\psi_j}\mathcal{L}_{\mathcal{T}^j}^a(\psi_j)$, 
		\State $\omega_j\leftarrow\omega_j-\mu\nabla_{\omega_j}\mathcal{L}_{\mathcal{T}^j}^c(\omega_j)$.
		\EndFor
		
		\State Empty the $j$th experience buffer.
		\EndFor
		\EndFor
		
		\State Update meta parameters with (\ref{eq13}): 
		\State $\psi\leftarrow\psi+\frac{\epsilon}{N_{\textrm{task}}}\sum_{j=1}^{N_{\textrm{task}}}(\psi_j-\psi)/\mu$, 
		\State $\omega\leftarrow\omega+\frac{\epsilon}{N_{\textrm{task}}}\sum_{j=1}^{N_{\textrm{task}}}(\omega_j-\omega)/\mu$.
		\EndFor
	\end{algorithmic}
\end{algorithm}
\begin{algorithm}[htbp]
	\caption{\large Meta-RL based Spectrum Sharing for V2X communication: Adaptation Stage}
	\label{alg2}
	\begin{algorithmic}[1]
		\Require
		meta parameters: $\psi_{\textrm{meta}}$ (actor), $\omega_{\textrm{meta}}$ (critic)\newline
		adaptation rate: $\mu$
		\Ensure
		policy: $\psi$ (actor), $\omega$ (critic)
		\State Initialize the policy with the meta parameters. 
		\State $\psi\leftarrow\psi_{\textrm{meta}}$,  $\omega\leftarrow\omega_{\textrm{meta}}$.
		\For{iteration $i\in[1, N_L]$}
		\State Collect experience data $\mathcal{D}$ with $\pi_{\psi}$.
		\State Store data to the experience buffer.
		\For{gradient step $s\in[1,N_U]$}
		\State Sample data from the experience buffer.
		\State Update parameters:
        \State $\psi\leftarrow\psi+\mu\nabla_{\psi}\mathcal{L}^a(\psi)$,
		\State $\omega\leftarrow\omega-\mu\nabla_{\omega}\mathcal{L}^c(\omega)$.
		\EndFor
        \State Empty the experience buffer.
		\EndFor
	\end{algorithmic}
\end{algorithm}

We adopt the PPO algorithm, which displays its state-of-the-art performance in \cite{30}, to train our agent. The PPO algorithm is an actor-critic method so the agent has an actor to select the action, whose parameters are denoted as $\psi$, and a critic to estimate the state value, whose parameter are denoted as $\omega$. For training the actor network, we need to maximize: 
\begin{equation}
\mathcal{L}^a(\psi)=\mathbb{E}_{t}[\min(r_t(\psi), \textrm{clip}(r_t(\psi),1-\epsilon,1+\epsilon))\hat{A}_t],
\label{eq8}
\end{equation}
where $r_t(\psi)=\frac{\pi_{\psi}(a_t|s_t)}{\pi_{\psi^{\textrm{old}}}(a_t|s_t)}$ denotes the probability ratio of importance sampling. The magnitude of the surrogate objective is limited by the clip function with clip parameter $\epsilon$, so the training process can stabilize. $\hat{A}_t$ is the generalized advantage estimation (GAE) function:
\begin{equation}
\hat{A}_t=\sum \limits_{i=0}^{T-t+1}(\gamma\lambda)^i\delta_{t+i},
\label{eq9}
\end{equation}
where $\lambda$ is the GAE parameter and $\gamma$ is the discount rate. $\delta_{t}$ is the temporal difference error:
\begin{equation}
\delta_{t}(\omega)=r_t+\gamma V_{\omega}(s_{t+1})-V_{\omega}(s_t),
\label{eq10}
\end{equation}
where $V_{\omega}(s_t)$ corresponds to the state value of $s_t$ and it is evaluated using the critic network, of which the weights are $\omega$. Besides, for training the critic network, we need to minimize:
\begin{equation}
\mathcal{L}^c(\omega)=\mathbb{E}_{t}[\delta_{t}^2(\omega)].
\label{eq11}
\end{equation}

\subsection{Design of the Outer Loop}
The outer loop amalgamates the update parameters of inner loops and updates the meta parameters, thus obtaining an optimal initialization for the policy network. The outer loop is designed to maximize the objective function below:
\begin{equation}
\mathcal{L}_{\textrm{meta}}(\psi)=\mathbb{E}_{\mathcal{T}^j\sim p(\mathcal{T})}[\mathcal{L}_{\mathcal{T}^j}^a(\psi_j)],
\label{eq12}
\end{equation}
with $\psi_j=\psi+\mu\nabla_{\psi}\mathcal{L}_{\mathcal{T}^j}^a(\psi)$, and $\mu$ is the step size. The parameters to be updated in the $j$th task are denoted as $\psi$, with $p(\mathcal{T})$ representing the distribution of learning tasks, and $\mathcal{T}^j$ indicating the $j$th task. Despite its desirable performance, MAML necessitates an extra backward process and supplementary trajectory sampling, resulting in high computational complexity. So, we have opted a comparable yet simplified meta-learning algorithm, Reptile \cite{17}. For Reptile, the meta parameters are updated as follows:
\begin{equation}
\psi\leftarrow\psi+\frac{\epsilon}{N_{\textrm{task}}}\sum_{j=1}^{N_{\textrm{task}}}(\psi_j-\psi)/\mu,
\label{eq13}
\end{equation}
with $N_{\textrm{task}}$ representing the batch size of sampled tasks. $\epsilon$ denotes the step size of the outer loop. Algorithm \ref{alg1} depicts the meta-training stage of our algorithm.
\subsection{The Adaptation Stage}
The meta parameters obtained in the meta-training stage will be used as the initialization of the neural networks for a new task. From the perspective of meta learning, these parameters enable the agent to learn how to extract useful information from a few experience data and help fine-tune its parameters for adaptation when confronting new tasks. The adaptation stage of our meta-RL based algorithm is shown in Algorithm \ref{alg2}.

The adaptation stage is similar to the inner loop of the meta-training stage. However, simulators are unnecessary in the adaptation stage, as the experience data is collected directly through interacting with the new vehicular environment with meta initialized policy. Although this policy may still select improper actions and cause some cost when collecting data, the probability is much lower than the random-initialized one. Besides, as meta parameters have extracted some essential information for solving these problems, much fewer interactions are required to obtain an efficient scheme for this new environment than learning from scratch, as will be shown in the experimental results. After the data collection, the PPO method described in Section ~\ref{inner} is used to update the parameters in the adaptation stage.

\section{SIMULATION RESULTS}
\label{simulation}
\subsection{Simulation Settings}
In this section, we present numerical results to validate the proposed meta-RL based resource allocation scheme for vehicular networks. Our simulator is custom-built and adheres to the evaluation methodology for the urban case as used in 3GPP TR 38.885 \cite{3}. Major simulation parameters are listed in Table \ref{tab1}.
\begin{table}[htbp]
\centering
\caption{Simulation Parameters}
\resizebox{\linewidth}{!}{
\begin{tabular}{c|c}\hline
Parameter & Value\\\hline
Number of vehicles $A$ & 4 \\
Carrier frequency & 2 GHz \\
Bandwidth& 4 MHz \\
BS antenna height & 25 m \\
BS antenna gain & 8 dBi \\
BS receiver noise figure & 5 dB\\
Vehicle antenna height & 1.5 m\\
Vehicle antenna gain & 3 dBi\\
Vehicle receiver noise figure & 9 dB\\
Noise power $\sigma^2$ & -114 dBm\\
Size of the area & 375 $\times$ 649 $\textrm{m}^2$\\
Power of V2I transmission  $P_a^{\textrm{V2I}}$ & 23 dBm\\
Time constraint of V2V transmission $T$ & 100 ms\\
Channel coherence time $\Delta T$ & 1 ms\\
V2V payload size $B$ & [1,2,$\dots$]$\times$1,060 bytes\\
Decorrelation distance of the V2I links & 50 m\\
Decorrelation distance of the V2V links & 10 m\\\hline
\end{tabular}}
\label{tab1}
\end{table}

Each V2V agent's actor and critic networks comprise three fully connected hidden layers, with 500, 250, and 120 neurons, respectively. The rectified linear unit (ReLU) is utilized as the activation function, and network parameters are updated using the Adam optimizer. The step size of the inner loop is set as, $\mu$=3e-4, and the step size of the outer loop is set as, $\epsilon$=1e-4. The discount rate is $\gamma$=0.99, and the GAE parameter is $\lambda$=0.95. The path loss model of the V2I links obeys the formulation $128.1+37.6\textrm{log}_{10}d$ (with $d$ in kilometers \cite{3}) while the V2V links follow the LOS in WINNER + B1 Manhattan \cite{3,31}. The shadowing distribution of both the V2I and V2V links is log-normal, and their shadowing standard deviations $\zeta$ are 8 dB and 3 dB, respectively. The slow fading components, i.e., the path loss and shadowing, update every 100 ms in the simulation.

\subsection{Convergence Analysis}\label{ca}
In this paper, we mainly consider changing five factors of the vehicular environment: the number of V2V links that each vehicle constructs with its neighboring vehicles, the size of V2V payload, the speed of the vehicles, and the fast fading types of these two links. For the fast fading type, we only consider the Ricean fading for simplicity, of which the Ricean factor is adjustable. We generate a meta training set comprising 243 tasks and a meta testing set of 16 tasks depending on this selection of factors. For each outer loop, a fixed number of tasks are randomly sampled from the task set for training, with the number as $N_{\textrm{task}}$=20. 
\begin{figure}[htbp]
    \centering
    \includegraphics[width=0.48\textwidth]{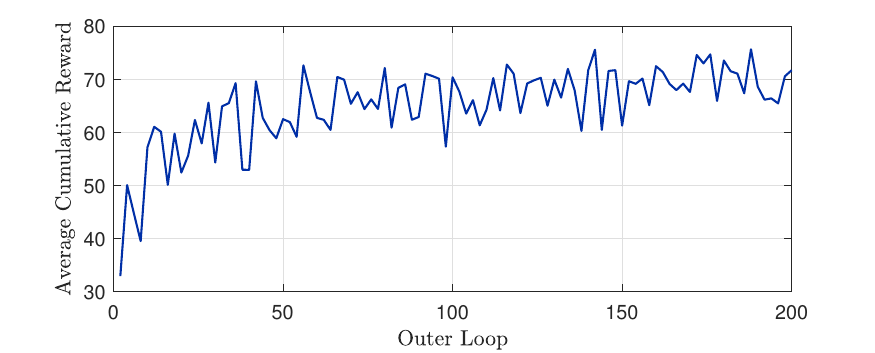}
    \caption{Convergence of meta training.}
	\label{fig2}
\end{figure}

The meta parameters are derived through $N_{\textrm{out}}$=200 outer loops in our experiment and the number of the inner loops is set as $N_{\textrm{in}}$=2. For each inner loop, 10 trajectories\inst{2}\let\thefootnote\relax\footnotetext{\inst{2}One trajectory refers to the experience data collected in one episode.} will be collected using the current policy $\pi_{\psi_j}$. Following the update of the meta parameters, we sample 10 tasks from the set for meta testing to conduct the evaluation of performance. It is noteworthy that we actually test the policy that is initialized with the updated meta parameters and rolls out 20 episodes (i.e., $N_L$=2) in the testing task for adaptation. 
\begin{figure}
	\centering
	\subfloat[V2I sum rate]{
		\includegraphics[width=0.48\textwidth]{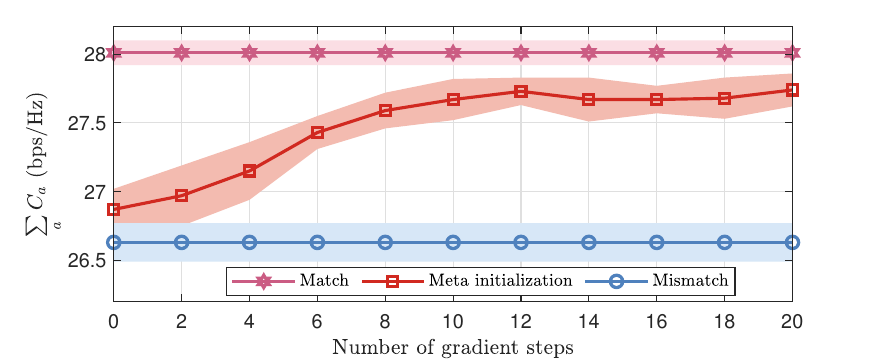}
	}
    \vspace{-0.1cm}
	\subfloat[V2V success transmission probability]{
		\includegraphics[width=0.48\textwidth]{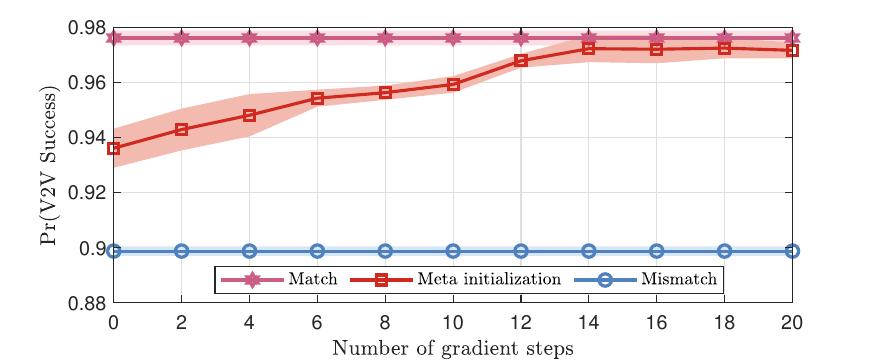} 
	}
	\caption{Performance with varying gradient steps in adaptation.}
    \label{fig3}
\end{figure}
We test the policy for 50 episodes per task and then calculate their averaged episodic cumulative reward. The result is presented in Fig. \ref{fig2}, which illustrates a rapid increase in the averaged episodic cumulative reward during the first 30 loops, with the reward value reaching an approximate convergence after about 150 loops.

We investigate the impact of meta parameters on the policy during adaption for a new task, as depicted in Fig. \ref{fig3}, which shows the performance of our scheme with the varying number of gradient descents. Here, only the number of gradient steps in adaptation changes, while the number of gradient steps for meta-training is fixed, as is the case in \cite{16}. Two baselines are considered:
\begin{itemize}
    \item Matched policy: a policy trained for 3,000 episodes in this new task.
    \item Mismatched policy: a policy trained for 3,000 episodes in a separate task, of which five main factors are different from this new task.
\end{itemize}

Fig. \ref{fig3} indicates that an increase in the number of gradient descents during adaptation corresponds to an elevation in both the sum rate of the V2I links and the success probability of  transmission of the V2V links. For a meta parameters initialized policy, about 14 gradient descents are needed in this new task to attain performance close to the matched policy.

According to our evaluation in several different environments, it usually needs 10 to 14 steps of gradient updates to achieve convergence. To decrease the number of gradient steps, we fix it as $N_U$=10 in all following experiments. We can see in the following experiments that 10 gradient steps are enough to achieve desirable performance.
\subsection{Performance evaluation}
To assess the influence of meta parameters on performance within a new environment, we conduct a comparison of the proposed methods with several baseline algorithms, as depicted in Fig. \ref{fig4} and Fig. \ref{fig5}, with regards to the augmentation of V2V payload sizes. The main difference between these two figures is that there are $N$=4 V2V links in Fig. \ref{fig4} while the number is set as $N$=8 in Fig. \ref{fig5}. We designate our proposed method as Meta initialization (20 episodes), indicating that we commence by initializing a policy with the meta parameters, followed by 20 episodes for adaptation in this new task. In addition to two baselines mentioned in Section \ref{ca}, \textcolor{blue}{three} other baselines are also considered, and we list them below:
\begin{figure}[ht!]
	\centering
	\subfloat[V2I sum rate]{
		\includegraphics[width=0.45\textwidth]{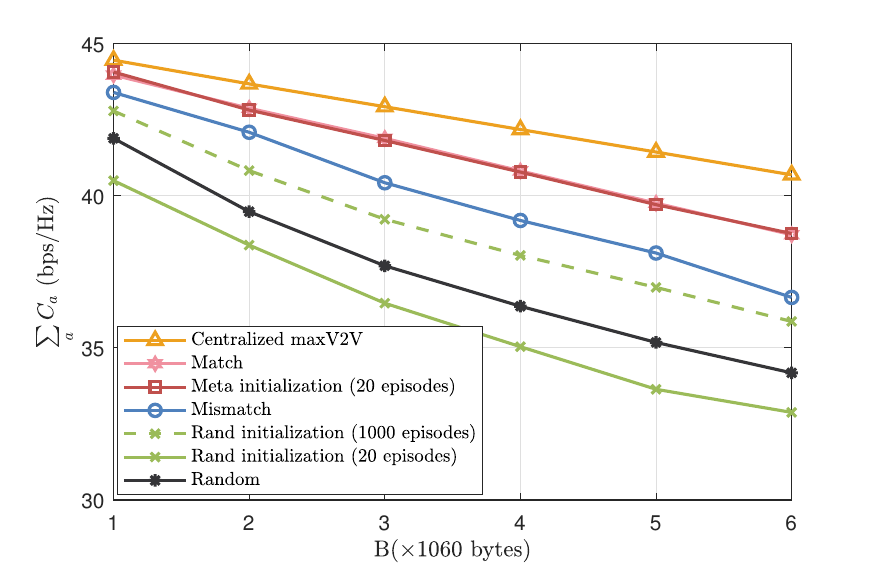} 
	}
    \vspace{-0.1cm}
	\subfloat[V2V success transmission probability]{
		\includegraphics[width=0.45\textwidth]{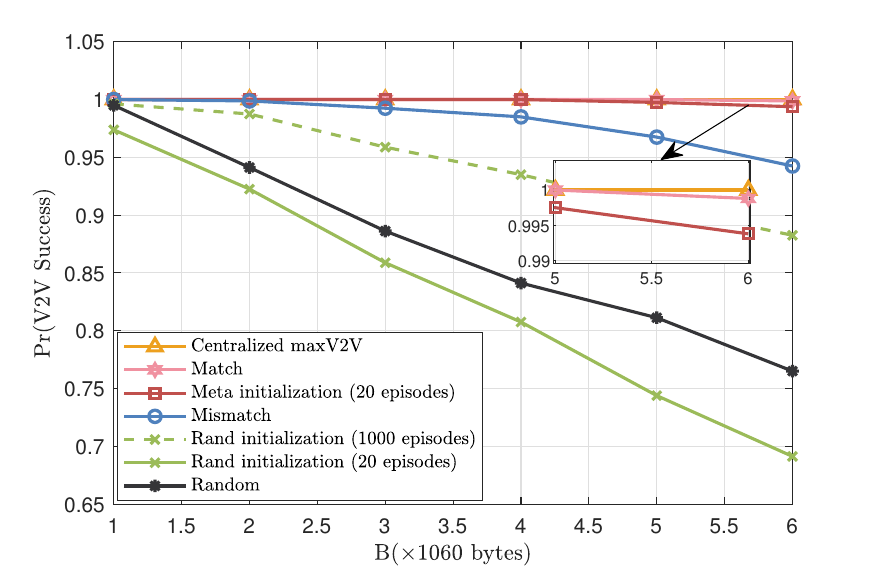} 
	}
	\caption{Performance with varying size of V2V payload $B$ ($N=4$).}
        \label{fig4}
\end{figure}
\begin{itemize}
\item  Rand initialization: a randomly initialized policy that is trained using PPO methods. The number of episodes for training is shown inside the bracket.
\item  Random: The spectrum to access and power for transmission are randomly selected by the agent.
\item  Centralized maxV2V \cite{6}: The sum V2V rate in each step is maximized by exhaustively searching the action space for all $N$ V2V links. This scheme requires accurate global CSI and a centralized implementation with extremely high complexity. We note that although this scheme is too idealistic to be used in practice, it can provide a meaningful performance upper bound for our analysis.
\end{itemize}
\begin{table*}[htbp]
\centering
\caption{The design of the meta training sets in Section \ref{sec44}}
\resizebox{\linewidth}{!}{
\begin{tabular}{c|c|c|c|c|c}\hline
 & Link & Velocity (Km/h) & Payload (bytes)& Ricean factor (V2I)& Ricean factor (V2V)\\\hline
72 tasks & 1, 3 &10, 20, 30&[2, 4, 6]$\times$1,060&10, 20&0, 6 \\
243 tasks & 1, 2, 3&10, 20, 30&[2, 4, 6]$\times$1,060&10, 15, 20&0, 3, 6 \\
432 tasks & 1, 2, 3&10, 20, 30&[2, 4, 6]$\times$1,060&10, 13.3, 16.7, 20&0, 2, 4, 6 \\\hline
\end{tabular}
}
\label{tab2}
\end{table*}

According to Fig. \ref{fig4}, the meta parameters initialized policy which subsequently rolled out 20 episodes for adaptation exhibits performance essentially equivalent to that of the matched policy, which has been trained in the same environment for 3000 episodes. 
\begin{figure}[ht!]
	\centering
	\subfloat[V2I sum rate]{
		\includegraphics[width=0.45\textwidth]{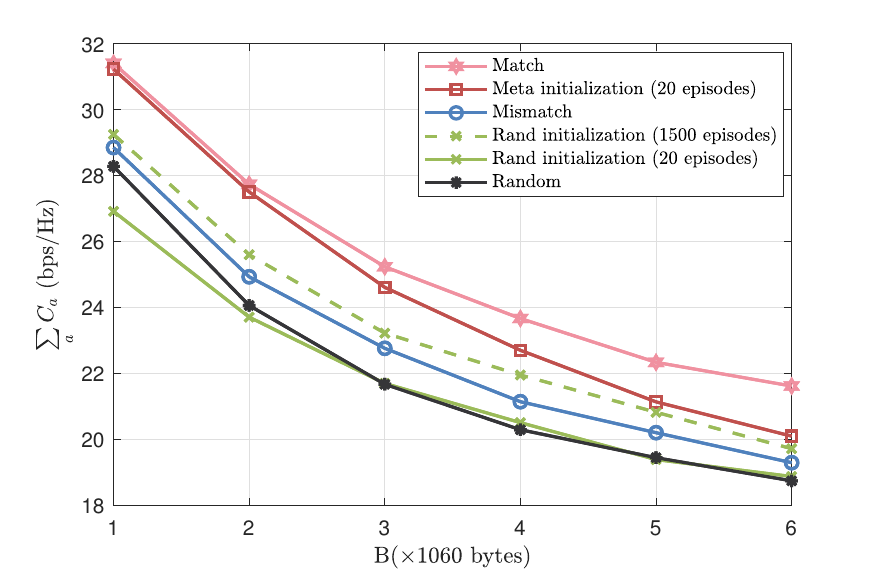} 
	}
    \vspace{-0.1cm}
	\subfloat[V2V success transmission probability]{
		\includegraphics[width=0.45\textwidth]{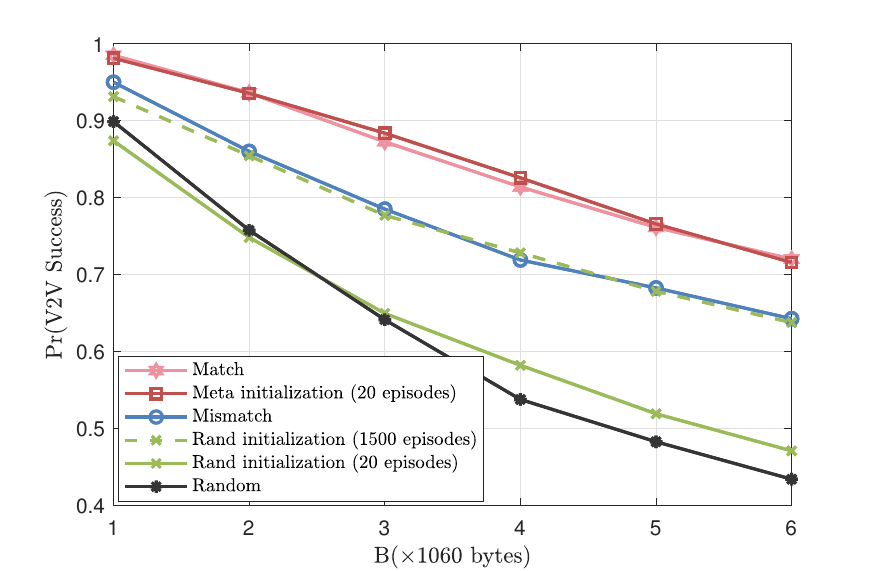} 
	}
	\caption{Performance with varying size of V2V payload $B$ ($N=8$).}
        \label{fig5}
\end{figure}
In addition, it performs measurably close to the centralized maxV2V scheme in that only about 4$\%$ of performance loss is observed for the V2I sum rate and no noticeable loss occurs in terms of V2V reliability. Moreover, our scheme only requires local CSI and the computational complexity is relatively low as it merely needs a forward pass. When compared with the matched policy, the mismatched policy suffers about $5\%$ degradation in both metrics in the worst case ($6\times1,060$ bytes of payload) due to the inherent difference in the training environments. In the case of a random-initialized policy, 20 episodes of adaptation are insufficient to effectively enable it to adapt to the new environment, of which performance is worse than the random baseline. From our perspective, the network is far from optimal or even functional with only 20 episodes of training. It will output bad actions from time to time, which is likely to be even worse than a random selection from the action set. Therefore, it can be concluded that the desirable performance of our method is mainly derived from the effective initialization, i.e., the meta parameters. For the random initialized policy, 1,000 episodes of training are necessary for it to attain significant performance improvement in the current setting.

When the number of V2V links in the system increases, interference becomes more severe, but we can find in Fig. \ref{fig5} that our algorithm is still effective. As the computation complexity of centralized maxV2V scheme grows exponentially with the number of agents, it is too time-consuming to obtain the results when $N$=8. So, this scheme is not included in Fig. \ref{fig5}. Although our method suffers about $6.8\%$ degradation compared with the matched policy in the sum rate of V2I links, it still outperforms the mismatched policy by about $4.95\%$ even in the worst case. While in the successful transmission probability of V2V links, our method attains the performance that is close to the matched policy. We also find that the mismatched policy suffers about $9\%$ degradation in both performance metrics, which is more severe than the situation in Fig. \ref{fig4}. The mismatched policy may suffer greater degradation if there are more V2V links in the system, which demonstrates the necessity of adopting meta-RL or other similar methods to tackle the problem of reality gap. We also note that it may require 1,500 episodes of training for the randomly initialized policy to obtain apparent improvement in this setting. Because the environment becomes more complex, the policy needs more experience data to acquire enough knowledge about it. 

Based on the results shown in Fig. \ref{fig4} and Fig. \ref{fig5}, the scalability of our meta-RL based method against V2V payload variation and the varying number of V2V links can be demonstrated. Our method achieves similar performance with the matched policy in these different situations, so we can conclude that the parameters suitable for initializing the policies can be found through our algorithm. Simultaneously, the required number of experience data in a new environment can be substantially decreased. In other words, an efficient policy for spectrum sharing can be obtained within fewer interactions, and this method is much faster than letting our policy learn from scratch in the new environment.  
\subsection{Impact of the number of training tasks}\label{sec44}
We further investigate how the total number of tasks in the meta training set affects the adaptation speed of policies in a new environment. We generate three meta training sets with 432 tasks, 243 tasks, and 72 tasks according to the varying values of the five factors. The corresponding values of five factors for each meta training set are listed in Table \ref{tab2}. For instance, a specific task in the meta training set of 243 tasks can be generated by selecting one of the optional values in the third row of Table \ref{tab2} for each of the five factors. As each factor has three values available, they can constitute $3^5$=243 tasks to form a complete set. Then we can obtain the meta parameters of each meta training set using Algorithm \ref{alg1}. It should be noted that the number of outer loops for each training set is different. For fairness, the times that each task is sampled in training should be kept roughly the same.

\begin{figure}[ht!]
	\centering
	\subfloat[V2I sum rate]{
		\includegraphics[width=0.45\textwidth]{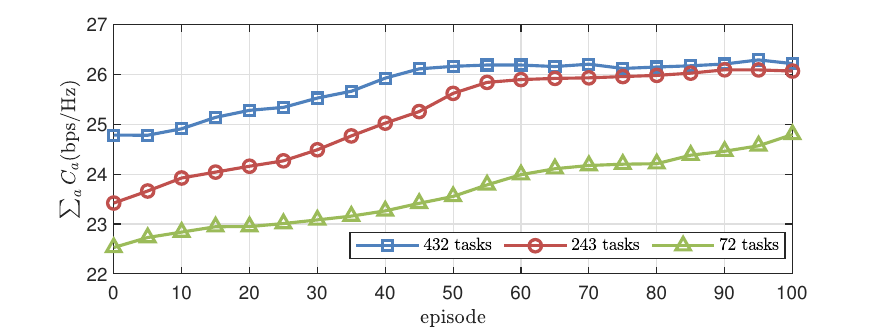} 
	}
    \vspace{-0.1cm}
	\subfloat[V2V success transmission probability]{
		\includegraphics[width=0.45\textwidth]{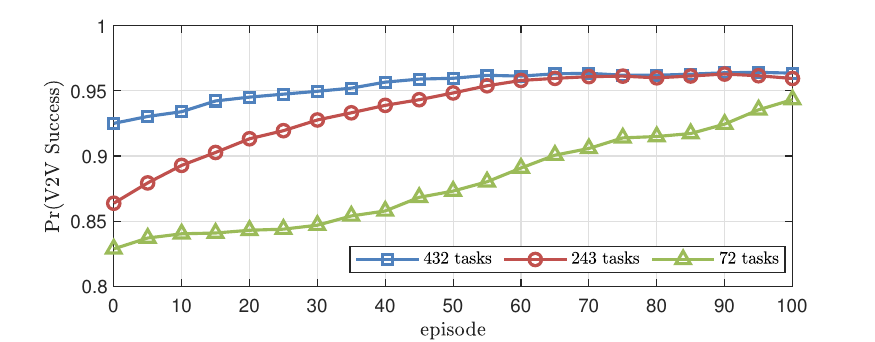} 
	}
	\caption{Impact of the number of tasks in the meta training set.}
        \label{fig6}
\end{figure}

We use these three trained meta parameters to initialize the policies and adapt them in an unseen environment for 100 episodes. We update these policies every 5 episodes during the adaptation, i.e., $N_L$=20. These policies are tested in the same environment for 100 episodes after each loop and this adaptation process is repeated 50 times to get the average performance. The results are shown in Fig. \ref{fig6}.

For both performance metrics, the policy curves initialized with the meta parameters trained with 432 tasks start at the highest level among these three curves and converge in approximately 45 episodes. Besides, the curves corresponding to 243 tasks need about 60 episodes to reach convergence. When there are only 72 tasks in the training set, the curves can not converge even in 100 episodes. Therefore, the total number of training tasks plays a vital role in improving the generalizability of the meta parameters. When the number of training tasks increases, more general information of the distribution of these tasks, $p(\mathcal{T})$, can be learned via meta-RL, leading to the meta parameters with better generalization. As the training tasks and the testing tasks are derived from the same distribution, meta-initialized agent has structured knowledge about the testing task so it can adapt to this new task within fewer interactions and exhibit faster convergence.
\subsection{Impact of the difference between testing and training environments}
\begin{figure}
	\centering
	\subfloat[V2I sum rate]{
		\includegraphics[width=0.45\textwidth]{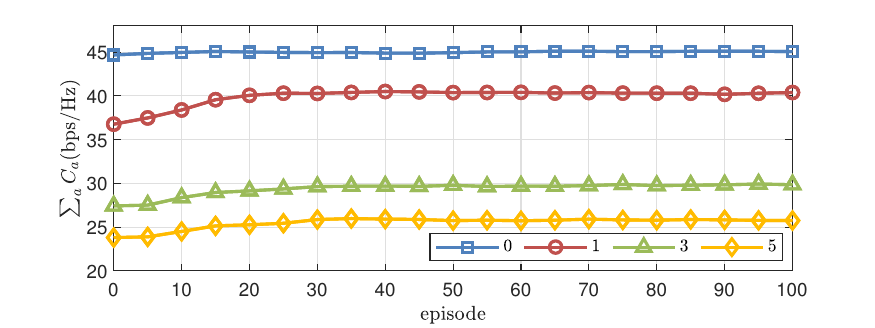} 
	}
    \vspace{-0.1cm}
	\subfloat[V2V success transmission probability]{
		\includegraphics[width=0.45\textwidth]{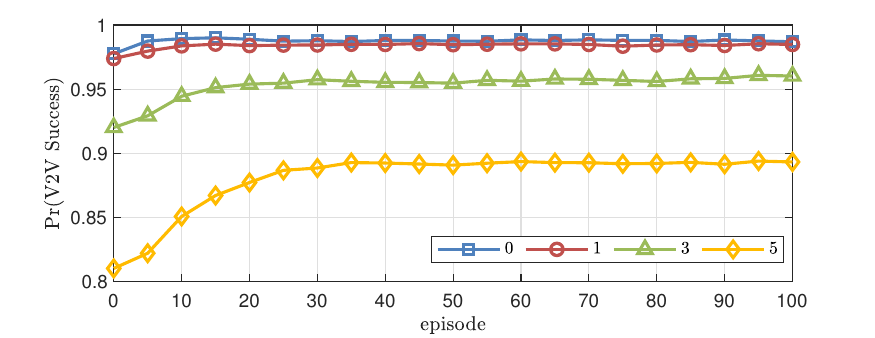} 
	}
	\caption{Impact of the difference between the testing and training environments.}
    \label{fig7}
\end{figure}

In this part, we examine how the meta trained policy adapts in an unseen environment depending on how much the new environment differs from those encountered in the training tasks. Here, the difference between environments is measured by how many of the five main factors of the environment are different.
% i.e., the number of neighboring V2V links, the V2V payload size, the average velocity of the vehicles, the fast fading types of the V2I and V2V links. 
We choose 4 testing tasks for experiments that differ from the training tasks in 0, 1, 3, and 5 factors, respectively. To distinguish the curves more clearly, we design the testing tasks based on the principle that the more different the testing task is, the worse the communication condition is. The experimental method is the same as the one of Fig. \ref{fig6}. The results are shown in Fig. \ref{fig7}. 

If all factors of the testing task are inconsistent with the training set, it needs about 40 episodes to converge in both performance metrics. In contrast, when facing the same task included in the training set, the policy only requires 5 or 10 episodes to achieve convergence. As the number of different factors in the testing task increases, it will require more episodes for the policy to converge. Therefore, we can demonstrate that the smaller the difference between testing and training tasks is, the faster the policy adapts, which is intuitively sensible. When the testing task differs from the training tasks in more factors, the agent may need more interactions to infer the new environment dynamics. When encountering the task once seen in the training, the agent already has enough knowledge about it, so fewer interactions are required.

\section{CONCLUSION}\label{conclusion}
In this paper, we propose an algorithm for fast spectrum sharing on the basis of meta-RL in vehicular networks. We give the problem formulation of spectrum assignment and power allocation at first and then turn it into a RL problem with the objective of simultaneously maximizing the sum rate of the V2I links and the probability of successful transmission of the V2V links. Rather than training policies for specific tasks, an algorithm based on meta-RL is leveraged to extract a good initialization for the policies. Numerical results demonstrate that our policy can adapt to a new task within a few episodes. Thus, we can achieve efficient spectrum sharing with a reduced frequency of interactions with the environment. Furthermore, we investigate the impact of the total number of tasks in the meta training set and the difference between the testing task and the training tasks on the performance of the policy. We demonstrate that when more tasks in the training set or the testing task differs from the training tasks to a lesser extent, the agent will adapt to the new task faster.

\bibliographystyle{ieeetr}
\bibliography{myref}

% \biographies

% \begin{CCJNLbiography}{photo.eps}{Bei Liu}
% received the B.S. and Ph.D. degrees from Beijing University of Posts and Telecommunications (BUPT) in 2001 and 2006, respectively. He is currently an Associate Professor in School of Electronic Engineering, BUPT. His current research interests include wireless communication theory and technology, wireless mesh networks.
% \end{CCJNLbiography}

% \begin{CCJNLbiography}{photo.eps}{Fei Zhang}
% received the B.S. degree in electronic science and technology from Beijing University of Posts and Telecommunications (BUPT) in 2017. He is currently working towards the master degree in School of Electronic Engineering, BUPT. His current research is signal processing technology for wireless communications.
% \end{CCJNLbiography}

\end{document}